# Acoustooptic Diffraction in Borate Crystals

[1]Martynyuk-Lototska I., [1]Dudok T., [1]Krupych O., [1]Adamiv V., [2]Smirnov Ye. and [1]Vlokh R.

Institute of Physical Optics, 23 Dragomanov St., 79005 Lviv, Ukraine
e-mail: vlokh@ifo.lviv.ua
Kyiv T.Shevchenko National University, 64 Volodymyrska St., 01033 Kyiv, Ukraine
e-mail: esmirnov@mail.univ.kiev.ua

## Abstract

The efficiency of acoustooptic (AO) diffraction in $\alpha$-BaB$_2$O$_4$ and Li$_2$B$_4$O$_7$ crystals is studied experimentally. The crystals are shown to be quite good AO materials. The efficiency of AO diffraction in $\alpha$-BaB$_2$O$_4$ reaches $\eta$=30% at the electric signal power of P=0.7W for the transverse acoustic wave and 15% at the power of P=0.56W for the longitudinal wave. The same parameter for Li$_2$B$_4$O$_7$ reaches $\eta$=21% at P=0,81W for the longitudinal acoustic wave.

## Introduction

As we have earlier shown [1,2], borate crystals, in particular $\beta$-BaB$_2$O$_4$ and $\alpha$-BaB$_2$O$_4$, should be efficient acoustooptic (AO) materials. According to the theoretical estimation, they possess high enough AO figures of merit (AOFM). For instance, in case of the optimised geometry for AO interaction, the AOFM can be as high as $M_2$=243.4×10$^{-15}$s$^3$/kg in $\alpha$-BaB$_2$O$_4$ crystals [3]. In other words, with respect to acoustooptics, at least some of the borate crystals may be compared with the best AO materials such as, e.g., TeO$_2$ crystals (the corresponding AOFM is equal to $M_2$=800×10$^{-15}$s$^3$/kg [4]). Besides, the borates possess higher resistance to optical radiation [5], when compare with TeO$_2$, and are also transparent in the UV spectrum range. So, Li$_2$B$_4$O$_7$ is transparent from 0.25μm up to 3.2μm [2]. Although the estimated AOFM of the lithium tetraborate, optimised with regard to the AO interaction geometry, is the lowest among $\alpha$-BaB$_2$O$_4$, $\beta$-BaB$_2$O$_4$ and Li$_2$B$_4$O$_7$ (namely, we have $M_2$=2.57×10$^{-15}$s$^3$/kg), the crystal manifests just the highest damage threshold. It is resistant to the optical radiation up to 32GW/cm$^2$ [5].

It is necessary to note that the AO interaction in borate crystals has been so far described theoretically and on the basis of experimental studies for the acoustic wave velocities, the refractive indices and the photoelastic coefficients. There is still no direct experimental data on the AO diffraction in borates. The present work is therefore devoted to this subject and deals with a particular case of Li$_2$B$_4$O$_7$.

## Experimental

Single crystals of $\alpha$-BaB$_2$O$_4$ and Li$_2$B$_4$O$_7$ have been grown with the Czochralski technique. Experimental investigations of AO diffraction in $\alpha$-BaB$_2$O$_4$ have been carried out for the two cases of AO interaction. In the first case, the transverse acoustic wave in α-BaB$_2$O$_4$ sample has been excited with piezoelectric LiNbO$_3$ transducer at the frequency of 26.5 MHz (the third harmonics of the transducer's resonance frequency, 8.8MHz). In the second case, the piezoelectric LiNbO$_3$ transducer has excited the longitudinal acoustic wave with the frequency of 100 MHz. Then the AO cell has been made, using the cold vacuum welding technique for connecting AO material with piezoelectric transducer. The AO cells based on Li$_2$B$_4$O$_7$ have been made for operating with the transverse acoustic wave at 50 MHz and the longitudinal wave at 100 MHz, with the aid of the same cold vacuum welding technique. In order to measure the dependence of AO diffraction efficiency in α-BaB$_2$O$_4$ and Li$_2$B$_4$O$_7$ crystals on the electric power applied to transducer, we have used He-Ne laser with the wavelength of 632.8 nm. The experimental setup is shown in Figure 1.

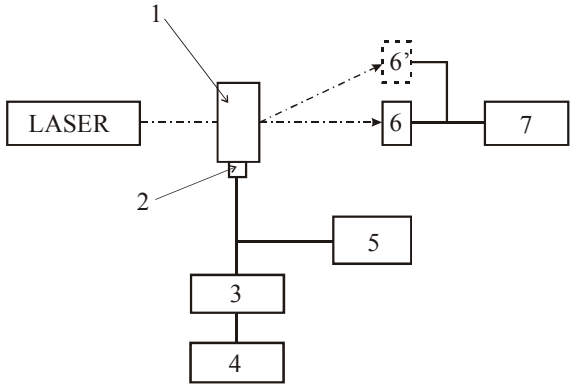

**Fig.1.** Experimental setup for investigations of AO diffraction: 1, AO crystal; 2, piezoelectric transducer; 3, amplifier; 4, high-frequency generator; 5, oscilloscope; 6, 6', photodetectors; 7, voltmeter.

The high-frequency electric signal from generator 4 has been amplified with amplifier 3 and then applied to piezoelectric transducer 2. The parameters of the electric signal at the transducer were controlled with oscilloscope 5. The incident optical radiation of He-Ne laser has propagated through AO crystal 1, in which the piezoelectric transducer 2 excites the transverse (or longitudinal) acoustic wave. The intensities of the incident light ($I_i$) and the diffraction maximum ($I_d$) have been detected with photodetector 6 and measured using voltmeter 7. The diffraction efficiency may be calculated with the relation

$$\eta = \frac{I_d}{I_i} \qquad (1)$$

The $I_d$ and $I_i$ values have been measured as functions of the driving electric power.

**Results and discussion**

The results of experimental studies for the AO diffraction efficiency in case of the interaction of optical wave with the transverse acoustic wave in α-BaB$_2$O$_4$ crystals are presented in Figure 2. Here the transverse acoustic wave at 26.5 MHz propagates along $x$ axis and is polarized along $z$ axis, while the light beam propagates along $y$ axis. The measured diffraction angle is equal to θ = 0,19°. The AO diffraction efficiency reaches η=30% in α-BaB$_2$O$_4$ crystal for the applied electric signal power of P=0,7W.

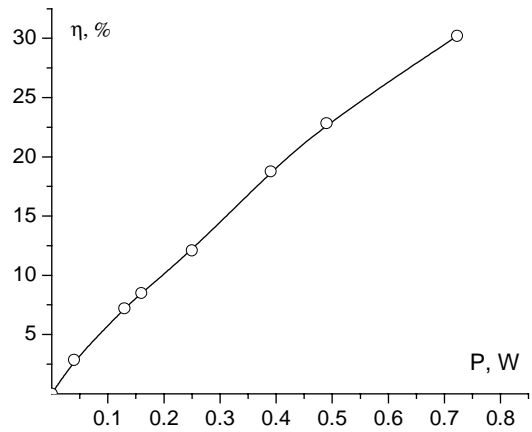

**Fig. 2.** Dependence of AO diffraction efficiency on the electric driving power for α-BaB$_2$O$_4$ crystals and the case of interaction with the transverse acoustic wave.

The AO diffraction in Li$_2$B$_4$O$_7$ has been experimentally studied for the interaction with the transverse acoustic wave (the frequency 50 MHz). Then the acoustic wave polarized along $z$ axis propagates in the direction of $x$ axis and the light beam is parallel to $y$ axis. In the latter case, the so-called Schaefer-Bergmann light scattering is observed (Figure 3a). The Schaefer-Bergmann scattering pattern in $Li_2B_4O_7$ corresponds to a cross section of the reciprocal ultrasonic velocity surface by xz plane (see Figure 3b).

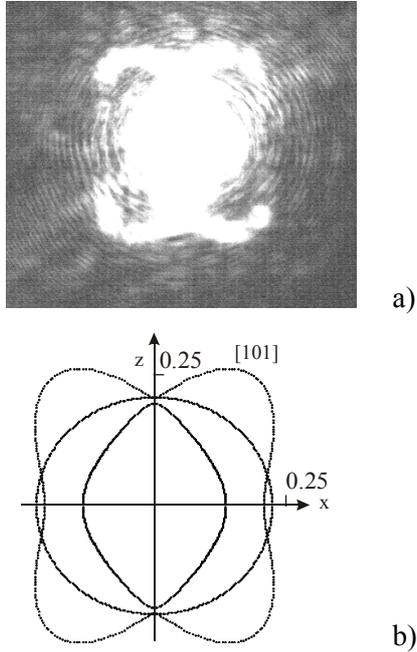

**Fig. 3.** Schaefer – Bergmann scattering pattern in $Li_2B_4O_7$ crystals appearing due to AO interaction of the acoustic wave excited along *x* axis and the incident optical beam propagated along *y* axis: (a) experimental results and (b) cross section of the surface of reciprocal acoustic wave velocity by *xz* plane.

The experimental results for the AO diffraction efficiency and the case of interaction with the longitudinal acoustic wave in $\alpha\text{-}BaB_2O_4$ and $Li_2B_4O_7$ crystals are presented in Figures 4 and 5.

Notice that in both AO cells, the longitudinal acoustic wave propagates along *z* axis and the light beam along *y* axis. The AO diffraction efficiency achieves $\eta=21\%$ in $Li_2B_4O_7$ crystals for the electric signal power of P=0.81W and 15% in $\alpha\text{-}BaB_2O_4$ crystals for P=0.56W. It is necessary to remark that the mentioned values do not represent a limit of some kind. At higher electric powers, it would be reasonable to expect a further increase in the diffraction efficiency.

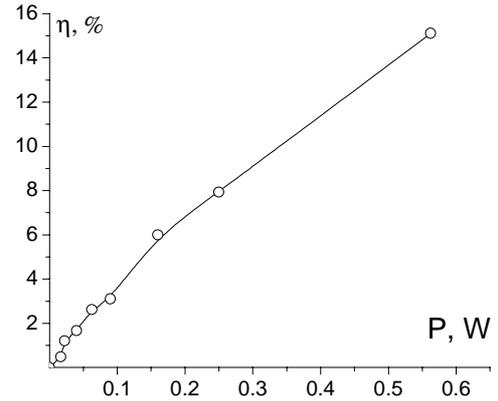

**Fig. 4.** Dependence of AO diffraction efficiency on the electric driving power for $\alpha\text{-}BaB_2O_4$ crystals and the case of interaction with the longitudinal acoustic wave.

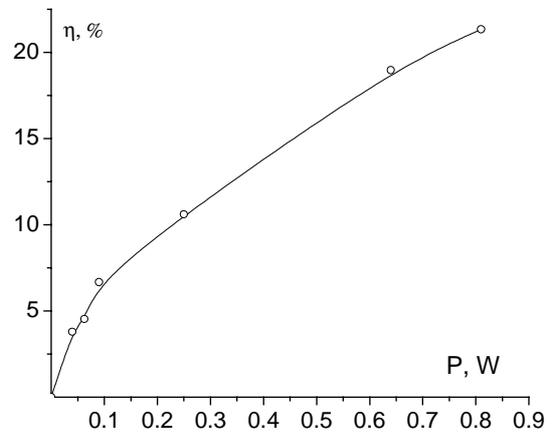

**Fig. 5.** Dependence of AO diffraction efficiency on the electric driving power for $Li_2B_4O_7$ crystals and the case of interaction with the longitudinal acoustic wave.

## Conclusion

We have shown experimentally that the diffraction efficiency of $\alpha\text{-}BaB_2O_4$ achieves the value of $\eta=30\%$ for the power P=0.7W of the applied electric signal and the transverse acoustic wave and 15% for the case of P=0.56W and the longitudinal wave. The diffraction efficiency of $Li_2B_4O_7$ may be as high as $\eta=21\%$, when the applied electric signal has the power of P=0.81W and the longitudinal acoustic wave is dealt with. At the same time, the values obtained above are in no case a limit of some kind. With increasing power of electric signal, it would be possible to get increasing diffraction efficiency.

Thus, the crystals under test may be successfully used as efficient AO materials.

**Acknowledgement**

The authors acknowledge financial support of the present work from the Scientific and Technology Centre of Ukraine under the Project N1712.